\documentstyle[aps]{revtex}
\begin{document}
\def\slash#1{#1 \hskip -0.5em / } 
\tighten
\def\jp{{J^P}}
\def\vp{v^\prime}
\def\ga{\gamma}
\def\g5{\gamma_5}
\def\beq{\begin{equation}}
\def\eeq{\end{equation}}
\def\mdJ{{\mu_1\dots\mu_J}}
\def\mdm{{\mu_1\dots\mu_k}}
\def\mdn{{\mu_1\dots\mu_n}}
\def\mdj{{\mu_1\dots\mu_j}}
\def\dex{D^{(j)}}
\def\ds{D^{(*)}}
\def\etmj{\eta^{\mdm}}

\begin{titlepage}
\begin{flushright}
JLAB-TH-97-35
\end{flushright}

\vspace{1.cm}
\begin{center}
\Large\bf Strong Decays Of Heavy Hadrons In HQET\end{center}
\vspace{0.5cm}
\begin{center}
W. Roberts
\\\vspace{3mm}
{\it  Department of Physics, Old Dominion University\\
Norfolk, VA 23529 USA \\ and \\
Thomas Jefferson National Accelerator Facility\\
12000 Jefferson Ave., Newport News, VA 23606, USA}
\end{center}

\vspace{1.5cm}
\begin{abstract}
We discuss the application of the tensor formalism of HQET to the strong decays
of heavy hadrons. We treat both meson and baryon decays, and note that all of
our results are in agreement with the `spin-counting' arguments of Isgur and
Wise. We briefly discuss the possible extension of the formalism to include
$1/m$ corrections.
\end{abstract}
\end{titlepage}

\section{Introduction}

In the past few years, the heavy quark effective theory (HQET) \cite{neubert,georgi,HQET} has enjoyed much
success in treating many aspects of the phenomenology of heavy hadrons \cite{baryons,LUKE}. The elegant
tensor formalism developed by Georgi, and used extensively by others, has
proven to be a very powerful tool for applications of HQET. There remains,
however, one area that is yet to be treated by this tensor formalism, namely
that of the strong decays of these hadrons. Clearly, since these decays, by
their very nature, involve non-perturbative QCD, we do not expect HQET to
allow us to calculate absolute decay rates. However, it will be useful
in examining ratios of decay rates.

The question of the ratios of decay rates has been addressed by Isgur and Wise \cite{IW}.
In their article, they noted that amplitudes for strong decays of heavy mesons
were proportional to sums of products of four Clebsch-Gordan coefficients that
arise from recoupling of angular momenta in the parent and daughter hadrons. In
fact, their result can be expressed slightly more compactly as a $6-J$ symbol.
Their argument makes use of the fact that the heavy quark is
a spectator in the decay of the heavy hadron, so that only the
light component of the hadron, the so called brown-muck, takes an active part
in the decay. 

At the hadronic level, a heavy hadron of total spin $S$ decays to
one of spin $S^\prime$, with a light hadron of total angular momentum $S_h$.
The daughter hadrons are in a relative $L$-wave, and conservation of angular
momentum gives
\begin{equation}
{\bf S_h+L=J_h, \,\,\, S_\ell^\prime+S_Q=S^\prime},
\end{equation}
with
\begin{equation}
{\bf S^\prime+J_h=S},
\end{equation}
where $S_\ell^\prime$ is the spin of the brown muck in the
daughter hadron, and $S_Q$ is the spin of the heavy quark.
This combination of angular momenta may be represented symbolically as
\begin{equation}
\left[\left[S_hL\right]_{J_h}\left[S_\ell^\prime
S_Q\right]_{S^\prime}\right]_S.\nonumber
\end{equation}
On the other hand, one can regard this process as proceeding entirely at the
level of the brown muck, since the heavy quark is a spectator in
the process, so that
\begin{equation}
{\bf S_\ell^\prime+J_h=S_\ell,\,\,\, S_\ell+S_Q=S}
\end{equation}
or
\begin{equation}
\left[\left[\left[S_hL\right]_{J_h}S_\ell^\prime\right]_{S_\ell}S_Q\right]_S.
\end{equation}

The overlap between these two `wave functions' or coupling schemes is
\begin{equation}
(-1)^{(S_Q+S_\ell^\prime+J_h+S)}\sqrt{(2S^\prime+1)(2S_\ell+1)}
\left\{\matrix{S_Q&S_\ell^\prime&S^\prime\cr
J_h&S&S_\ell\cr}\right\}.
\end{equation}
This object is proportional to the strong
matrix element, and the constants of proportionality are the same for the four
decays that are possible between two different multiplets. That is, there exist
a single set of proportionality constants for the four decays
\begin{equation}
\left(S_\ell\pm 1/2\right) \to \left(S_\ell^\prime\pm 1/2\right) +h.
\end{equation}

In this article, we
show how to use the tensor method to obtain the same information in a manner
that we find somewhat more compact than the `spin-counting' method. In
addition, the specific forms of the amplitudes can be obtained using this
formalism. Furthermore, the full power of the tensor formalism
may then be brought to bear on these processes. For instance, it should be
possible to treat the $1/m_Q$ corrections, as well as the radiative ones, to
the decay amplitudes.

\section{Tensor Formalism}

\subsection{General Formalism}

In general, we are interested in a matrix element of the form
\begin{equation}
{\cal M}=\left<{\cal X}(p) {\cal H}^\prime_Q(v^\prime)\left|{\cal O}_s\right| {\cal
H}_Q (v)\right>,
\end{equation}
where ${\cal X}$ is a light hadron, ${\cal H}^\prime_Q$ and  ${\cal H}_Q$ are
heavy hadrons, and ${\cal O}_s$ is the operator responsible for the strong
decay. The problem in trying to say anything useful about this lies in the fact
that ${\cal O}_s$ is, in general, a complicated object that is full of
non-perturbative QCD dynamics, and about which we know very little.
In general, this operator will involve all of the sub-structure of the hadron
in a non-trivial way.

Despite this difficulty, we do know that ${\cal O}_s$ must be a Lorentz scalar, as well as 
flavor singlet in all flavors of quarks. In particular, it is flavor singlet in the heavy quark.
For the purposes of this discussion, and without any loss of generality, we can write ${\cal O}_s$ as
\begin{equation}\label{operator}
{\cal O}_s=\sum_ia_i\overline{Q}\Gamma_i Q L_i,
\end{equation}
where $\Gamma_i$ is a general Dirac matrix, and is one of $1,\,\gamma_\mu,\, \sigma_{\mu\nu},
\,\gamma_\mu\gamma_5,\,\gamma_5$, and $L_i$ contains all of the dynamics involving the brown muck. The $a_i$ are
unknown constants. We have written ${\cal O}_s$ in this form in order to explicitly display
the heavy quark part. Each
$L_i$ has the same Lorentz structure as the corresponding $\Gamma_i$, to ensure that
${\cal O}_s$ is a Lorentz scalar.  Note that the $L_i$ are, in principle, many body operators, as the 
structure of the brown muck is expected to be complicated. While it may be tempting to associate the
$\overline{Q}\gamma_\mu Q$ term of ${\cal O}_s$, for instance, with `one-gluon physics', we refrain from making 
such identifications. This is because we view eqn. (\ref{operator}) simply as a way of parametrizing our 
ignorance of strong interaction dynamics, and make no interpretations of the physics that could lead to each term.

Since we can represent the heavy hadrons as ${\rm heavy \,part}\times{\rm light
\,part}$, such as in
the Falk representation \cite{falk},
the matrix element of each term in ${\cal O}_s$ factorizes. This means that although
the interactions involving the light component are complicated, we can absorb
these into a set of unknown form factors (as has been done for weak decays) or unknown coupling
constants (for strong decays). All that is left for us to deal with are the
heavy quark components, which we know how to treat. Furthermore, we also know
how to include corrections due to the finite mass of the heavy quark. However, there are still five terms in
${\cal O}_s$.

What helps us further in our treatment of the strong process is the fact that, at
leading order in HQET ({\it i. e.}, in the limit that the mass of the quark $Q$
goes to infinity), the heavy quark will act as a spectator in the decay. In particular, its spin indices are
unaffected by the decay (this is the same physics contained in the spin-coupling scheme described above). Thus,
the only possible form that can contribute is the scalar contribution, $\Gamma_i=1$. Beyond leading order we
would expect other terms to contribute. This also means
that the light part of the matrix element is simplified, as the operator concerned is a Lorentz-scalar.
This identification, that only $\Gamma_i=1$ above can contribute at leading order, is the key to applying the 
tensor formalism to these decays. All else is
now relatively simple, as we know how to `calculate' matrix elements for any
arbitrary $\Gamma$, as well as how to include various kinds of corrections.

We close this section by noting that the coupling constants to which we have
alluded, which are essentially the matrix elements of the light part of the
decay operator, depend only on the brown muck, and are therefore independent of
the mass of the heavy quark. Thus, for instance, the same set of coupling
constants would be valid for decays of hadrons containing $c$ quarks and for
hadrons containing $b$ quarks.

\subsection{Kinematic Questions}

HQET, in conjunction with chiral perturbation theory (ChPT), has been applied to the strong single (and
double) pion decays of heavy hadrons \cite{goity,falk1,falk2}.  In that treatment, the requirement that the pion momentum be
small, combined with the `infinite' masses of the parent and daughter hadrons, leads to the fact
that the velocity of the heavy daughter hadron is the same as that of the parent. This is in
contrast with the weak decays of these states, in which the heavy quark can receive a large
momentum change from the emitted virtual $W$: the velocities of the parent and daughter hadrons are
different. In the HQET/ChPT formalism for strong decays, there are therefore two independent 
kinematic variables, $v$, the velocity of the heavy hadrons, and $p$, the momentum of the pion.

In the present formalism, we want to be able to treat the decays of a heavy hadron to another heavy
hadron, with the emission of a `light' hadron that may be any of the infinite tower of excited
states. Thus, if the light daughter is sufficiently excited, it can provide the large impulse
required to bring about a velocity change in the heavy hadrons. In fact, note that if we were to
consider decays of charmed mesons to the $a_2$, say, the mass of the light daughter hadron is
already a sizable fraction (70\%, in the case of decays to the ground state charmed mesons) of the
mass of the daughter hadron. It therefore appears necessary to make use of full momentum
conservation through
\begin{equation}
m_{D^{(j)}}v=m_{D}\vp+p,
\end{equation}
where $v$ is the velocity of the parent, $\vp$ is that of the heavy daughter, and $p$ is the
momentum of the light daughter.

We could therefore use two of the three quantities $v$, $\vp$ and $p$ as independent kinematic
variables. In this case, since the velocity of the heavy hadron has changed, we are apparently 
implicitly including $1/m_Q$ (and higher) corrections that are of a purely kinematic nature, and
which have no effect on the spin symmetry of HQET. We note, however, that it has become customary
to use the physical momentum of the light hadron in examining these decays. This amounts, in
essence, to a second choice of kinematics. 

We note, however, that since the splittings between states of the heavy spectra are independent of the mass of
the heavy quark, so too is $p$, the momentum of the light hadron produced in the decay. This means that $v-\vp$
must scale as $1/m_Q$. For the purposes of `power counting', it is therefore more convenient to use $p$ as one of
the variables, instead of $v-\vp$. We have therefore chosen $p$ and $v$ as our kinematic variables. We close by noting 
that the choice of kinematics will probably become more important for the consideration of
$1/m_Q$ corrections.

\section{Meson Decays}

The starting point for this discussion is the representation of the heavy meson
states. For concreteness, let us examine decays of excited $D^{**}$ mesons to
ground state $D^{(*)}$ mesons. For these, we use the representations
constructed by Falk \cite{falk}.

An excited $D$ meson with total angular momentum $J$ will, in general,
be represented by an object linear in a polarization tensor, $\eta^{\mu_1\dots\mu_J}(v)$. 
This polarization
tensor is symmetric, transverse and traceless. The latter two properties are expressed by
\beq
v_{\mu_1}\eta^{\mu_1\dots\mu_J}(v)=0,\,\, g_{\mu_1\mu_2}\eta^{\mu_1\dots\mu_J}(v)=0.
\eeq
For a state consisting of a heavy quark $Q$ and a light component with the quantum numbers of
an antiquark, the specific representation of any particular state will depend on the angular 
momentum $j$ of the light component (antiquark) of the state. It is thus more convenient to 
refer to $j$ than to $J$, since there will be a degenerate doublet of states with $J=j\pm 1/2$.
The full details of the representations can be found in Falk's article \cite{falk}.

We illustrate the tensor method for calculating strong decay matrix elements by examining two 
specific sets of decays: the
generalization to other cases should be obvious. We begin by looking at decays
involving single pions, so that we are interested in the matrix element
\begin{equation}
{\cal M}=\left<D(v^\prime)\pi(p)\left|\overline{c} c\right|\dex(v)\right>.
\end{equation}
As identified in the previous section, we are taking the heavy quark operator
responsible for the decay as $\overline{c} c$ (and a light scalar operator is understood as 
multiplying $\overline{c} c$). Note, too, that we are explicitly not
using the `chiral limit' of soft pions, as we allow $p$ to be large. In other
words, the velocity in the heavy daughter baryon is not the same as in the
parent, and momentum is conserved explicitly through
\begin{equation}
m_{D^{(j)}}v=m_{D}\vp+p.
\end{equation}
In terms of the trace formalism, the matrix element of interest is
\begin{equation}
\left<D(v^\prime)\pi(p)\left|\overline{c}c\right|\dex(v)\right>=\sqrt{M_DM_{\dex}}
{\rm Tr}\left[\Pi(p){\cal A}_{\mdm}\overline{{\cal D}}(\vp){\cal M}_{\dex}^{\mdm}(v)\right],
\end{equation}
where ${\cal D}(\vp)$ is the matrix representation of the meson $D$. The matrix ${\cal A}_{\mdm}$ can 
only have the form
\begin{equation}
{\cal A}_{\mdm}=p_{\mu_1}\dots p_{\mu_k},
\end{equation}
while the matrix $\Pi(p)$ must represent the final state pion. The simplest,
non-redundant form allowable is 
\begin{equation}
\Pi(p)=a\g5,
\end{equation}
where the constant $a$ is independent of the mass of the heavy quark, by virtue
of our chosen normalization.
One could also include a term in $\slash{p}$, but this is redundant.
Thus, the matrix element is
\begin{equation}
\left<D(v^\prime)\pi(p)\left|\overline{c}c\right|\dex(v)\right>=a\sqrt{M_DM_{\dex}}
{\rm Tr}\left[\g5\overline{{\cal D}}(\vp){\cal M}_{\dex}^{\mdm}(v)\right] p_{\mu_1}\dots p_{\mu_k}.
\end{equation}
Due to the spin symmetry of HQET, the decays to the corresponding vector meson
$D^*$, are also described by the same coupling constant $a$, and the
corresponding matrix element is
\begin{equation}
\left<D^*(v^\prime,\varepsilon)\pi(p)\left|\overline{c}c\right|\dex(v)\right>=a\sqrt{M_{D^*}M_{\dex}}
{\rm Tr}\left[\g5\overline{{\cal D}^*}(\vp){\cal M}_{\dex}^{\mdm}(v)\right] p_{\mu_1}\dots p_{\mu_k}.
\end{equation}
Thus, these four decays are all described in terms of a single, unknown,
nonperturbative constant $a$. 

We now turn to meson decays that are not as
simple. We limit our discussion to decays involving light vector mesons
($\rho$, for instance), but the generalization to light hadrons of arbitrary
spin should be clear. The matrix element for such a decay (still considering
decays to the ground state heavy doublet) is
\begin{eqnarray}\label{rho1}
\left<\ds(v^\prime)\rho(p,\epsilon)\left|\overline{c}c\right|\dex(v)\right>=&&\sqrt{M_{\ds}M_{\dex}}\nonumber\\
&&\times{\rm Tr}\left[{\cal R}(p)\epsilon^{*\nu}{\cal A}_{\nu\mdm}\overline{{\cal M}}_{\ds}(\vp) {\cal M}_{\dex}^{\mdm}(v)\right].
\end{eqnarray}
The most general form for the matrix ${\cal A}_{\nu\mdm}$ is
\begin{equation}
{\cal A}_{\nu\mdm}=p_{\mu_1}\dots p_{\mu_{k-1}}\left[av_\nu p_{\mu_k}+
b\ga_\nu p_{\mu_k}+cg_{\nu\mu_k}\right],
\end{equation}
while ${\cal R}(p)=1$ is the most general, non-redundant form that represents the $\rho$ meson (the
polarization vector of the $\rho$ appears explicitly in eqn. (\ref{rho1})). For a decay in which the parent belongs to
one of the $\left(0^-,1^-\right)$ or $\left(0^+,1^+\right)$ multiplets, the term $c g_{\nu\mu_k}$ is
absent, as there are then no indices on the matrix representation of the parent hadron.

We close this section with a brief discussion of the relationship between the formalism presented here and
that of the combined heavy quark effective theory and chiral perturbation theory, for decays involving
pions. It is clear that the two approaches are attacking the problem from somewhat different starting
points. What we have presented here does not require any constraint on the momentum of the pion produced in
the decay, and so may be considered as a formalism that automatically includes all the powers of pion
momentum that would arise in a chiral expansion. Note, for
instance, that for the $D$-wave decays of the $\left(1^+,2^+\right)$, we would write down the form that
we have written above, while in the chiral approach, operators with two powers of the pion momentum must
be explicitly constructed \cite{falk1,falk2}. Finally, we note that it is essentially trivial
to include two or more pions in any of these decays using this formalism.

\section{Baryon Decays}

The case of baryon decays may best be subdivided into two separate classes. The
first set of decays that we will treat are those in which the heavy daughter
hadron is a baryon (such as $\Lambda_b^*\to\Lambda_b\rho$), while in the second class, the heavy daughter hadron will
be a meson (such as $\Lambda_b^*\to pB$).

\subsection{Heavy Daughter Baryons\label{baryons2}}

As with the meson decays, our starting point is the representation of the baryon states. We will simply
borrow the representations constructed by Falk. We note, however, that we must divide our baryons into two
classes, those with `natural' parity, and those with `unnatural' parity. This description is determined by
the spin and parity of the brown muck, denoted $j^P$. If $P=(-1)^j$, the baryon is a natural one, while
if $P=(-1)^{(j+1)}$, the baryon is unnatural. The need for this division into natural and unnatural baryons 
will become clear shortly.

Consider the decay $\Sigma_b^{(j)}\to\Lambda_b\pi$, which is described by the matrix element
\begin{equation} \label{amdj}
\left<\Lambda_b(v^\prime)\pi(p)\left|\overline{b}b\right|\Sigma_b^{(j)}(v)\right>=\overline{u}(\vp) R_{\mdj}(v)
{\cal A}^{\mdj},
\end{equation}
where the spinor-tensor $R_{\mdj}(v)$ represents both states of the doublet, and ${\cal A}^{\mdj}$
contains all of the strong interaction dynamics. Since there is a pion in the decay, and assuming that the
$\Lambda_b$ is the ground state, then the quantity ${\cal A}^{\mdj}$ must be a pseudo-tensor if the
$\Sigma_b^{(j)}$ is a natural baryon, or a tensor if it is unnatural. The forms that can be constructed in
the two cases are quite different. Let us now examine some more specific examples.

Consider the decay $\Sigma_b^{(*)}\to\Lambda_b\pi$, where $\Sigma_b^{(*)}$ belongs to the
$\left(1/2^+,3/2^+\right)$ doublet. The matrix element is
\begin{equation}
\left<\Lambda_b(v^\prime)\pi(p)\left|\overline{b}b\right|\Sigma_b^{(*)}(v)\right>=\overline{u}(\vp) R_{\mu}(v)
r^{\mu},
\end{equation}
with $r^\mu$ a vector, which can only have the form
\begin{equation}
r_\mu=ap_\mu.
\end{equation}
On the other hand, if the $\Sigma_b^{(*)}$ belongs to the $\left(1/2^-,3/2^-\right)$ doublet, then $r^\mu$
would be a pseudo-vector, which cannot be constructed from the quantities we have at our disposal. Thus
\begin{equation}
\left<\Lambda_b(v^\prime)\pi(p)\left|\overline{b}b\right|\Sigma_b^{(1/2^-,3/2^-)}(v)\right>=0.
\end{equation}
The generalization of this to parent hadrons of higher spin is easy, since then ${\cal A}^{\mdj}$ of eqn.
(\ref{amdj}) becomes
\begin{equation}
{\cal A}_{\mdj}=ap_{\mu_1}\dots p_{\mu_j}
\end{equation}
for parents of unnatural parity, or
\begin{equation}
{\cal A}_{\mdj}=0
\end{equation}
for parents of natural parity.

In the case of the heavy baryons of natural parity, the amplitudes for decays to the ground state
with the emission of a single pion vanish at leading order, and should first be non-zero at order
$1/m_Q$. Thus, if these states have no other open channels into which they can decay, they should be
quite narrow.

For decays to final states that are not the ground state, such as to the $(1/2^+,3/2^+)$ multiplet,
the decay amplitude is
\begin{equation}
\left<\Lambda_b^{(1/2^+,3/2^+)}(v^\prime)\pi(p)\left|\overline{b}b\right|\Sigma_b^{(j)}(v)\right>=
\overline{R}_\nu(\vp) R_{\mdj}(v)
{\cal A}^{\nu\mdj}.
\end{equation}
If the parent hadron has natural parity, then ${\cal A}^{\nu\mdj}$ is a tensor (because the daughter
has unnatural parity), and takes the form
\begin{equation}
{\cal A}_{\nu\mdj}=p_{\mu_1}\dots p_{\mu_{j-1}}\left[ag_{\nu\mu_j}+bp_\nu p_
{\mu_j}\right].
\end{equation}
For parents of unnatural parity,
\begin{equation}
{\cal A}_{\nu\mdj}=ap_{\mu_1}\dots p_{\mu_{j-1}}\varepsilon_{\nu\mu_j\alpha\beta}
v^\alpha p^\beta.
\end{equation}

As final examples of the application of the formalism to this kind of decay we consider decays to 
$\rho$ mesons. The matrix element for decays to the ground state is
\begin{equation}
\left<\Lambda_b(v^\prime)\rho(p,\epsilon)\left|\overline{b}b\right|\Sigma_b^{(j)}(v)\right>=
\overline{u}(\vp) R_{\mdj}(v)\epsilon_\nu^*
{\cal A}^{\nu\mdj}.
\end{equation}
If the parent has natural parity, then ${\cal A}^{\nu\mdj}$ is a tensor and takes the form
\begin{equation}
{\cal A}_{\nu\mdj}=p_{\mu_1}\dots p_{\mu_{j-1}}\left[ag_{\nu\mu_j}+bv_\nu p_
{\mu_j}\right].
\end{equation}
For a parent of unnatural parity,
\begin{equation}
{\cal A}_{\nu\mdj}=ap_{\mu_1}\dots p_{\mu_{j-1}}\varepsilon_{\nu\mu_j\alpha\beta}
v^\alpha p^\beta.
\end{equation}

For decays to the $(1/2^+,3/2^+)$ multiplet, we obtain
\begin{equation}
\left<\Lambda_b^{(1/2^+,3/2^+)}(v^\prime)\rho(p,\epsilon)\left|\overline{b}b\right|\Sigma_b^{(j)}(v)\right>=
\overline{R}_\nu(\vp) R_{\mdj}(v)\epsilon_\alpha^*
{\cal A}^{\nu\alpha\mdj}.
\end{equation}
For parents of unnatural parity,
\begin{eqnarray}
{\cal A}_{\nu\alpha{\mdj}}&=&p_{\mu_1}\dots p_{\mu_{j-2}} \left\{ ag_{\nu\mu_j}g_{\alpha\mu_{j-1}}\right.\nonumber\\
&+&\left.p_\beta\left[bg_{\nu\alpha}g_{\beta\mu_j}+cg_{\nu\mu_j}g_{\alpha\beta}+dg_{\alpha\mu_j}g_{\nu\beta}
+eg_{\beta\mu_j}p_\alpha v_\nu\right]\right\},
\end{eqnarray}
while for parents of natural parity,
\begin{eqnarray}
{\cal
A}_{\nu\alpha\mdj}&=&p_{\mu_1}\dots p_{\mu_{j-1}}\left\{a \varepsilon_{\mu_j\nu\alpha\beta}
p^\beta\right.\nonumber\\
&+&\left.v^\beta p^\rho\left[bp^\alpha\varepsilon_{\mu_j\nu\beta\rho}+
cv^\nu\varepsilon_{\mu_j\alpha\beta\rho}+dp^{\mu_j}\varepsilon_{\nu\alpha\beta\rho}\right]
\right\}.
\end{eqnarray}

\subsection{Light Daughter Baryons}

For these decays, as with the decays to heavy baryons, it will again be useful
to divide the light baryons into natural and unnatural baryons, with a slight
modification of the definition. A light baryon of spin $J$ and parity $P$ is
considered to be natural if $P=(-1)^{J-1/2}$, unnatural if $P=(-1)^{J+1/2}$. In
addition, for states with spin greater than 1/2, we employ the generalized
Rarita-Schwinger fields $u_{\mdj}(p)$, which satisfy
\begin{eqnarray}
\slash{p}u_{\mdn}(p)&=&mu_{\mdn}(p),\nonumber\\
\gamma^{\mu_1}u_{\mdn}(p)&=&0,\nonumber\\
p^{\mu_1}u_{\mdn}(p)&=&0,
\end{eqnarray}
where $n=J-1/2$, and we remind the reader that this object is symmetric in all of its Lorentz
indices.

We first consider the decays $\Lambda_b\to B^{(*)} N$, where $N$ is the ground
state nucleon, and the parent represents any of the states that belong to
one of the $1/2^+$ singlets. The amplitude for the decay is
\begin{equation}
\left<B^{(*)}(v^\prime)N(p)\left|\overline{b}b\right|\Lambda_b(v)\right>=
\overline{u}(p) {\cal A} \overline{{\cal M}}_{B^{(*)}}(\vp) u(v),
\end{equation}
where ${\cal A}$ is the most general scalar matrix that can be constructed, and ${\cal M}_{B^{(*)}}(\vp)$
is the matrix representing the $B^{(*)}$ states.
Without loss of generality, we can choose ${\cal A}=a$, a constant.

We can generalize this for the decay of any excited state, such as
$\Lambda_b^{(j)}\to B^{(*)}N$. The amplitude is
\begin{equation}
\left<B^{(*)}(v^\prime)N(p)\left|\overline{b}b\right|\Lambda_b^{(j)}(v)\right>=
\overline{u}(p) {\cal A}^{\mdj}\overline{{\cal M}}_{B^{(*)}}(\vp) R_{\mdj}(v).
\end{equation}
For parents of natural parity, ${\cal A}^{\mdj}$ is a tensor, and takes the form
\begin{equation}
{\cal A}^{\mdj}=p^{\mu_1}\dots p^{\mu_{j-1}}\left[ap^{\mu_j}
+b \gamma^{\mu_j}\right].
\end{equation}
For parents of unnatural parity, ${\cal A}^{\mdj}$ is a pseudotensor, which we
can construct very easily (in this case) by using the tensor of the preceding
equation, and multiplying it by $\gamma_5$. Thus, for parents of unnatural
parity,
\begin{equation}
{\cal A}^{\mdj}=p^{\mu_1}\dots p^{\mu_{j-1}}\left[ap^{\mu_j}
+b \gamma^{\mu_j}\right]\gamma_5.
\end{equation}

Since the use of the $\gamma_5$ allows us to go from pseudotensor to tensor for
these decays, in what follows we will discuss only the decays of parent baryons
with natural parity.

The last set of decays we consider are $\Lambda_b^{(j)}\to
B^{(*)}\Delta$. For the amplitude, we write
\begin{equation}
\left<B^{(*)}(v^\prime)\Delta(p)\left|\overline{b}b\right|\Lambda_b^{(j)}(v)\right>=
\overline{u}_\nu(p) {\cal A}^{\nu\mdj} \overline{{\cal M}}_{B^{(*)}}(\vp)R_{\mdj}(v).
\end{equation}
Since the $\Delta$ has unnatural parity, ${\cal A}^{\nu\mdj}$ must be a pseudotensor
(for parents of natural parity), and takes the form
\begin{eqnarray}
{\cal A}_{\nu\mdj}&=&\left\{p_{\mu_1}\dots p_{\mu_{j-1}}\left[av_\nu
\gamma_{\mu_j}+bg_{\nu\mu_j}+cv_\nu p_{\mu_j}\right]\right.\nonumber\\
&+&\left.dp_{\mu_1}\dots p_{\mu_{j-2}}g_{\nu\mu_{j-1}}\gamma_{\mu_j}\right\}
\gamma_5.
\end{eqnarray}

\section{Discussion and Conclusion}

In the previous sections, we have outlined how the tensor formalism of HQET may be used to examine the strong
decays of heavy hadrons. There remain a few points of the formalism that warrant some discussion. First, note
that we have not presented any decay rates. Nevertheless, we have examined many cases for these decays, and
have found that the ratios of decay rates predicted by Isgur and Wise are indeed obtained.

We have not treated the decays of heavy mesons to a heavy baryon and a light antibaryon. However, the formalism for
these decays is very similar to that of the last subsection.

There is one subtlety involved in some of the matrix elements we have shown. Let's examine the case of the
$(1/2^-,3/2^-) \to 1/2^+\rho$, where the $1/2^+$ is the heavy baryon singlet. In this case, the $6-J$ symbol
becomes $$\left\{\matrix{1/2&0&1/2\cr J_h&S&1\cr}\right\},$$ which implies that $J_h$ can only have the value
1, regardless of the value of $S$. However, in our formalism, we have used two independent coupling
constants, implying two independent amplitudes. The resolution of this apparent contradiction lies in
realizing that for this decay, $J_h=1$ can be constructed in two different ways, with $L=0$ or $L=2$ (for the
$\rho$, $S_h=1$). Thus, there are indeed two independent amplitudes, corresponding to the two independent
partial waves, but the ratio of the two $L=0$ decay amplitudes is the same as that of the two 
$L=2$ decay amplitudes.

In their formalism, Isgur and Wise have pointed out that the total widths for the decays of the two members 
of a heavy spin multiplet to the two members of another multiplet are identical. In principle, we can obtain
this general result in the present formalism, but a proof is beyond the scope of the present article, and is
left for possible future work. We note, however, that for all of the cases we have examined explicitly, the
sum rule has been found to be valid, as expected.

In subsection \ref{baryons2}, we saw that there were some decays that vanished exactly at this order in the
$1/m$ expansion. For such amplitudes, the $1/m$ `corrections' are therefore the leading terms, and we believe
that these corrections should be studied. In addition, it is important to examine the $1/m$ corrections for
the non-vanishing amplitudes, as these may lead to large departures from the leading order predictions. This
has been done by Falk \cite{falk1,falk2} for the D-wave decays of the $(1^+,2^+)$ $D^{**}$ mesons to the ground states, in the
framework of the combined HQET and chiral perturbation theory. It is of some interest to see the kind of
contributions that can arise in the present formalism. In particular, an $S$-wave component is
expected for one of these decays.

The coupling constants we have introduced are all independent of the mass of the heavy quark present in the
parent and daughter hadrons. By virtue of the heavy flavor symmetry, these coupling constants are therefore
valid both for charm and beauty decays. Thus, knowing some charmed decay rates, we could predict the
corresponding beauty decay rates.
Alternatively, we could attempt to extend this formalism down to strange hadrons, treating the $s$ quark as
heavy, to glean some information about what to expect in charm. In this case one would certainly expect 
$1/m$ corrections to be very important.

Finally, we close on a very speculative note. The key to the formalism presented herein was the
identification of the heavy quark current that plays a role in the decay. For the strong decays, this current
was identified as being the unit Dirac matrix. It is possible that this idea can be extended to,
for example, electromagnetic processes of heavy mesons. In the decay $D^*\to D\gamma$, for instance, it is 
expected that the photon will couple both to the heavy quark and to the brown muck. We know what to do in the
first case, but not in the second. In the second case, however, we may still be able to use the idea that for
this part of the current, the heavy quark is a spectator, so that the heavy quark current is again unity, and
one is then left with the matrix elements of the light current, which may be parametrized in some way. This has
been done, to some extent, by a number of authors \cite{radiative}.
Whether this approach leads to any further development remains to be seen.

\section*{Acknowledgement}

Thanks go to J. Goity and N. Isgur for discussions and comments, and for reading the manuscript. Thanks also go to Institut des
Sciences Nucl\'eaires, Grenoble, France, where part of this work was done. This work was supported by the
National Science Foundation through grant \# PHY 9457892, and by the Department of Energy, through contracts
DE-AC05-84ER40150 and DE-FG05-94ER40832.


\vskip 0.25in

\begin{references}
\bibitem{neubert}{See, for example, M. Neubert, Phys. Rep. {\bf 245C} (1994)
259, and references therein.}
\bibitem{georgi}{H. Georgi, Phys. Lett. {\bf B240} (1990) 447.}
\bibitem{HQET}{N. Isgur and M. Wise, Phys. Lett. {\bf B232} (1989) 113; Phys. Lett. {\bf B237} 
(1990) 527; B. Grinstein, Nucl. Phys. {\bf B339} (1990) 253; A. Falk, H. Georgi, B. Grinstein and 
M. Wise, Nucl. Phys. {\bf B343} (1990) 1; 
A. Falk and B. Grinstein, Phys. Lett. {\bf B247} (1990) 406; T. Mannel, W. Roberts and Z. Ryzak, 
Nucl. Phys. {\bf B368} (1992) 204.}
\bibitem{baryons}{N. Isgur and M. B. Wise, Nucl. Phys. {\bf B348} (1991) 276; H. Georgi, 
Nucl. Phys. {\bf B348} (1991) 293; T. Mannel, W. Roberts and Z. Ryzak, Nucl. Phys. {\bf B355} 
(1991) 38.}
\bibitem{LUKE}{M. Luke, Phys. Lett. {\bf B252} (1990) 447; H. Georgi, B. Grinstein and M. B. 
Wise, Phys. Lett. {\bf B252} (1990) 456; C. G. Boyd and D. E. Brahm, Phys. Lett. {\bf B254} 
(1991) 468; A. Falk, B. Grinstein and M. Luke, Nucl. Phys. {\bf B357} (1991) 185.}
\bibitem{IW}{N. Isgur and M. Wise, Phys. Rev. Lett. {\bf 66} (1991) 1130.}
\bibitem{falk}{A. Falk, Nucl. Phys. {\bf B378} (1992) 79.}
\bibitem{goity}{J. L. Goity and W. Roberts, Phys. Rev. {\bf D51} (1995) 3459.}
\bibitem{falk1}{A. Falk and T. Mehen, Phys. Rev. {\bf D53} (1996) 231.}
\bibitem{falk2}{A. Falk, Preprint JMU-TIPAC-96014, unpublished.}
\bibitem{radiative}{P. Colangelo, F. De Fazio, G. Nardulli, Phys. Lett. {\bf B316} (1993) 555;
J. G. Korner, D. Pirjol, K. Schilcher, Phys. Rev. {\bf D47} (1993) 3955;
    H. - Y. Cheng {\it et al.}, Phys. Rev. {\bf D47} (1993) 1030.}

\end{references}
\end{document}